# Claude for Excel is Transformative for Basic Spreadsheet Modeling and Analysis

Thomas A. Grossman, Min-San Serena Hsu

University of San Francisco

tagrossman@usfca.edu, mhsu12@dons.usfca.edu

## ABSTRACT

*We used the Claude Excel plug-in to build and analyze spreadsheet models for five semi-structured problem statements (including a test-piece teaching case with data in multiple exhibits) and evaluated performance on modeling (what is to be coded), spreadsheet engineering (quality of coding), analysis (algorithms applied to the spreadsheet), and insight generation. Claude's ability to create a high-quality spreadsheet model from a semi-structured problem statement is astoundingly excellent. Given very simple prompts, Claude generates extensive, accurate sensitivity and scenario analyses and reasonable plain-English insights. Claude is Solver-aware and builds optimization models without specific prompting. Analyst productivity gains will likely be high. Business school quant courses will have to change. The way we develop human expertise will have to change.*

## 1 INTRODUCTION

It is widely-understood that artificial intelligence (AI) is affecting all aspects of business. There is currently a robust debate in the literature, popular press, informed social media, and uninformed social media on the impact of AI "large language models" (LLMs) on software engineering productivity.

A few minutes with the Gemini AI tool suggests that AI amplifies the gap between great developers, who use AI to automate boilerplate and syntax so they can focus on higher-level architecting, and less experienced developers who blindly trust AI, producing bloated code. There may be a transition from "skill as production" (knowing how to write the code) to "skill as judgment" (knowing how to read, edit, and orchestrate the code). These are early days and it is too soon to know how this will play out. However, it is clear that for at least some developers working in some coding shops, productivity is skyrocketing.

Spreadsheets are a powerful fourth-generation programming language [Grossman, Mehrotra & Özlük, 2007] ubiquitous in business and especially in finance. The lead author teaches business analytics and related topics in an AACSB-accredited business school. In the foundational course, we aim to teach students to be effective business analysts, and we use the Excel spreadsheet as the coding platform.

This paper explores the effectiveness of a leading commercial AI Excel plug-in (Claude for Excel) on the foundational business analytics skills in Excel.

### 1.1 Overview

We start in section 2 with a literature review (a challenge in a fast-moving field). We then (section 3) lay out the foundational topics for business analytics based on the Business Analytics Lifecycle. In section 4 we discuss Anthropic's Claude for Excel product. Section 5 examines Claude's effectiveness in creating a spreadsheet model from a problem statement. We consider (section 6) Claude's analysis of a spreadsheet model (sensitivity/scenario analysis, optimization, and generating VBA code). We make observations in section 7. In section 8 we share conclusions and topics for future research.

### 1.2 Contribution

This paper frames AI tools for spreadsheet modeling and analysis in terms of the Business Analysis Lifecycle. We examine the effectiveness of a leading spreadsheet AI tool (Claude for Excel) for Modeling, Spreadsheet Engineering, Analysis, and Insight Generation using a corpus of models and problem statements from MBA teaching.

Given a spreadsheet model, Claude performs sensitivity analysis and optimization with aplomb (including the advanced tactic of constraint status cells), and generates sophisticated analytical insights. It works around Microsoft's API limitations on Data Table by compressing a model into one or more cells.

For creating a spreadsheet model from a problem statement, we find that different AI models have very different performance, and low-tier models create errors. The high-tier models have extraordinary effectiveness and generate spreadsheet models that are comparable to those of a strong MBA student.

For an experienced spreadsheet analyst, Claude likely provides a sizeable productivity boost even after taking time to check the results. For a beginner analyst who cannot evaluate the quality and accuracy of the AI output, the value of the AI is problematic. In one instance Claude changed a data value rending the results technically inaccurate, so careful verification must be undertaken before results can be trusted.

Claude for Excel is transformative for spreadsheet modeling and analysis. Business school quant courses need an overhaul. The question of how we train future experts looms large.

## 2 LITERATURE REVIEW

For simplicity, we use "AI" to mean commercial LLMs. AI is moving quickly, as is AI in spreadsheets. There is a robust competitive marketplace in AI products intended to support users in building spreadsheet models and using them to extract insight. [Thorne, 2025] provides a useful overview that highlights the weaknesses of LLMs at the time of writing.

Things are moving so fast that a formal literature review seems pointless. For example [Grossman, Chen & Datuashvili, 2025], which pre-dated the release of Claude for Excel, found that ChatGPT was very limited in spreadsheet modeling. That finding has been overtaken by recently-released LLM products. We note that [Grossman et al 2025] was optimistic about the potential for AI in spreadsheet development, although the response of the audience at EuSpRIG 2025 was pointedly more skeptical; it will be fascinating to see how things evolve in the coming years.

Therefore, the reader interested in the current collective understanding of AI in spreadsheets is advised to use an AI-based internet search tool such as Google's Gemini, ChatGPT, or Claude for up-to-date information.

## 3 FOUNDATIONAL TOPICS FOR BUSINESS ANALYTICS IN EXCEL

The Excel business analyst starts with a business situation, creates a model (a set of mathematical equations), implements the model as a spreadsheet computer program, analyzes the model to identify quantitative model insights, interprets the model insights to generate managerial insights, and ultimately communicates the findings to influence the business situation. This is illustrated in the Business Analysis Lifecycle (Figure 1).

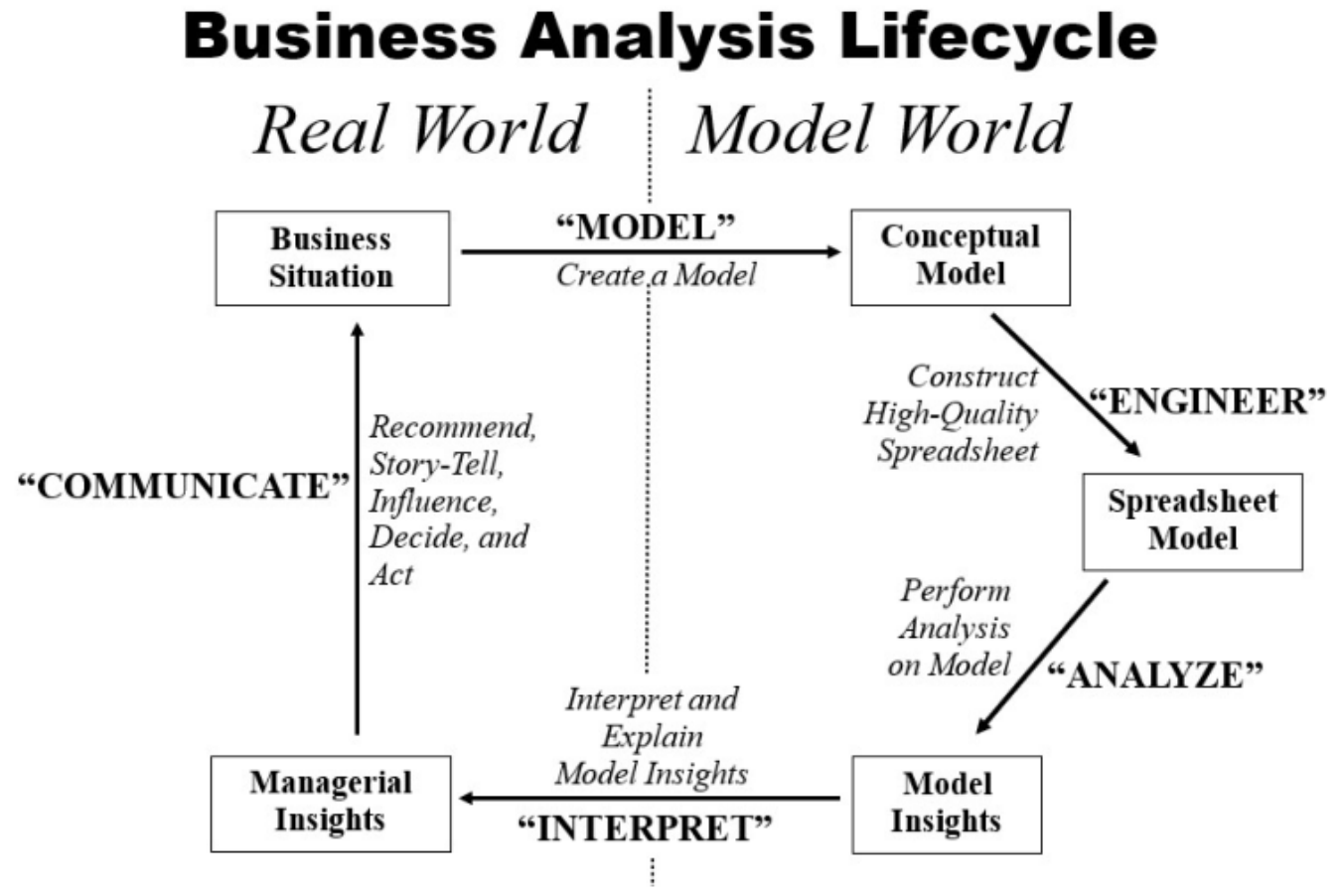

*Figure 1, The Business Analysis Lifecycle*

An initial version of the lifecycle first appeared in [Powell & Baker, 2007]. Iterations can be found in [Grossman et al, 2008] and [Grossman, Mehrotra & Sidaoui, 2016]. (The reader interested in the oft-overlooked "Model" step is referred to [Powell, 2001; Powell & Baker, 2004].) We believe that anyone building spreadsheets for business, including financial modeling, necessarily engages with all the steps in the lifecycle.

Within the "Analyze" step are a number of basic analytical techniques [Powell & Baker, 2007; Grossman, 2008] including Sensitivity Analysis (simple what-if analysis, one-parameter sensitivity analysis, tornado chart, two-parameter sensitivity analysis, multi-parameter sensitivity analysis which is implemented as scenario analysis) and Backsolving (finding input value(s) that generate a desired output value, called "breakeven analysis" in accounting, "inverse analysis" in engineering, and "goal-seeking" in Excel). Note that One- and Two parameter sensitivity analysis in Excel is automated using Data Table. There are also advanced analytics techniques including optimization (Excel Solver), risk analysis, and many other specialized techniques.

This paper considers model-driven analysis, where the intellectual effort is on articulating and coding a model that captures the essential aspects of a business situation, and then applying analytical techniques to obtain insight. We are not considering data-driven analysis, where the focus is on extracting insight and meaning from a dataset.

We report the results of experiments using AI that start with a business situation and lead to a spreadsheet model, and experiments that analyze a spreadsheet model using several basic and advanced analytical techniques.

## 4 CLAUDE FOR EXCEL

We used Anthropic's "Claude for Excel" product (henceforth referred to as "Claude"). Claude for Excel adds a "Claude" button to the Home ribbon. This button opens a Claude window within Excel where the user can select which Claude model to use, upload files, and write prompts. Although we did not do it for this paper, Claude allows the user to upload many documents for reference and context.

Claude for Excel is part of "Claude Pro" which at time of writing can be purchased for $200 per year. Claude for Excel was released in beta in October 2025. It was released to pro-tier subscribers in January 2026. A major update was released in March 2026. More updates can be expected (indeed, a new model called *Fable* was released while this paper was receiving final edits).

At the time of our experiments in Spring 2026, Claude had three model tiers called *Haiku*, *Sonnet*, and *Opus*. (And each has its own numbered versions.) Claude summarizes them as follows: "Haiku is the fastest and lightest option — built for everyday, quick requests like simple summaries, basic formulas, or straightforward data lookups, and it's the most efficient with your usage allowance. Sonnet is the recommended default for most work: it brings strong reasoning to coding, analysis, writing, and complex problem-solving, and notably handles spreadsheet tasks well — making it the best starting point if you're unsure which to pick. Opus is the heavy-duty reasoning specialist, best reserved for genuinely complex, multi-step analytical tasks that need deep thinking — in an Excel context, that might mean intricate financial modeling, multi-variable analysis, or building sophisticated logic you need thoroughly reasoned through. The practical rule of thumb: use Haiku for quick, repetitive tasks; Sonnet for most analytical and formula work; and Opus only when a task is complex enough to justify the extra reasoning depth it brings." Our experience is consistent with this summary.

## 5 CREATING A SPREADSHEET MODEL

Creating a spreadsheet model for a business situation (see Figure 1) requires two distinct activities: *Modeling* (create a conceptual model, which is a specification of a set of mathematical equations that can be represented verbally, algebraically, or in software including Python and Excel); and *Engineering* (create a spreadsheet computer program that implements the model). Spreadsheet analysts often perform both activities simultaneously, but experience teaches that treating them separately is helpful for efficiency, quality control, and training. Modeling is the realm of the domain expert or SME, and spreadsheet engineering is the realm of the coding expert.

We examined five business situations that were written up as "problem statements". Each problem statement was a PDF or DOCX containing a few paragraphs or pages of prose, with numeric values embedded in the prose and sometimes provided as a separate table(s). We uploaded to Claude the problem statement and a simple prompt. We used five problem statements, described in Table 1.

| Problem Statement | # Words | # Data Values | Type | Source |
|---|---|---|---|---|
| Wall Bid | 130 | 11 (all in-text) | Research Problem | [Panko & Sprague, 1998] |
| Asilomar Company | 205 | 19 (7 in-text + 12 in the sales/cost table) | Exercise | T. Grossman |
| Direct Response | 169 | 9 (all in-text) | Exercise | V. Mehrotra |
| Baker Products | 259 | 16 (all in-text) | Mini-Case | S. G. Powell |
| Buckeye Power & Light | 1420 | 87 in exhibits); 14 more in-text | Teaching Case | [Frey & Carraway, 1991] |

*Table 1, Description of Problem Statements*

The prompts were simple, for example, "Do a complete business analysis, present everything in excel with formulas." or "Based on the given files, do a complete business analysis, present everything in excel with formulas and provide the best recommendations". It is possible that details of the prompts affect Claude's performance, but the quality of output was so high that we did not explore this hypothesis.

Claude responds by presenting its reasoning (or at least it claims to do so); creating a model represented in Python code; and then converting the Python into Excel code. (As an aside, we hypothesize that PY files are preferable to XLSX files for extracting patterns from a corpus of models.)

### 5.1 Claude's Performance is Excellent

For each problem statement, Claude quickly (within a few minutes) produced an Excel file containing cell formulas (Appendix 1 contains an example of high-quality Claude output for the relatively simple Wall Bid problem statement). Claude sometimes produced multiple worksheets, and some level of analysis (discussed in section 6), and even plain-English summaries of findings. The Claude-produced spreadsheets all had good designs with informative row- and column-labels, and effective formatting including colors and boxes. These spreadsheets would be satisfactory as student-submitted homework.

Evaluating the correctness of a spreadsheet is inherently challenging. We are fortunate that the problem statements have been used by many spreadsheet developers (in some cases by many years of students) and hence we possess spreadsheet models for each problem statement that we are confident are fully correct. For each problem statement, we compared the outputs of our model(s) to those of a Claude model for both the base case, and after changing some model inputs. We also perform a brief review of the Claude code to confirm that cell formulas are present and appropriate. We did not perform a formal code inspection but given the modest size of these models that seems excessive. We state with confidence that Claude's spreadsheet models were correct. However (section 7.3) Claude did change a data value.

Based on these five problem statements, Claude can produce high-quality, accurate, reusable spreadsheet models that are suitable for professional use.

### 5.2 Is Claude Cheating?

It is hypothesized that AI does better if it works on a problem that it has already seen. (For example, the Wall problem statement, extracted from [Panko & Sprague 1998], is widely used in classroom teaching and we and our students have fed it to LLMs dozens of times.) We have not formally tested this possibility. However, the Asilomar Company exercise has never been provided to students and we believe it has never previously been fed to an LLM, and we found that Claude produced a high-quality spreadsheet model and analysis.

## 6 ANALYZING A SPREADSHEET MODEL

We performed several experiments that gave Claude a spreadsheet model, and prompted Claude to perform basic analysis.

### 6.1 Sensitivity Analysis (Data Table in Excel) Has a Strong Workaround

Sensitivity Analysis is the act of varying one or more model inputs and recording one or more model outputs. Sensitivity is likely the most common analytical technique. In Excel, it is automated using the Data Table feature. An Excel Data Table is a powerful feature, not least because it is "live" allowing other inputs to be varied in real time.

When prompted to "perform a complete sensitivity analysis", Claude helpfully explained that it does not have access to Data Table due to a lack of an API, and executed a workaround where it condensed the model into a single cell, made multiple copies of that cell down the rows of the spreadsheet, and generated output that looks the same as Data Table output. For our tests, accuracy was perfect.

In our experience, students value the power and flexibility of Data Table, but dislike its fiddly, non-intuitive interface. Claude's approach seems beneficial, in that it delivers the goods without having to learn and remember how to configure and launch Data Table. However, unlike Data Table, the table breaks if the model logic is modified.

### 6.2 Scenario Analysis (Sensitivity Analysis with more than two variables)

We teach a scenario analysis technique that uses a lookup technique to feed the data values for each scenario to the model, and uses a Data Table to summarize the results. We avoid the terrible Excel Scenario Manager [Grossman & Özlük, 2009].

Claude simply brute-forces the scenarios. It compresses the model's cell formulas into one or more cells, and copies these cells into a row for each scenario. For our one test, accuracy was perfect. However, unlike the lookup approach, the scenario outputs break if the model logic is modified.

### 6.3 Optimization Analysis (Solver in Excel) is Excellent

Optimization is an advanced analytical technique that is often (incorrectly) referred to as "linear programming". All business school management science textbooks (e.g., [Powell & Baker 2007]) have multiple chapters on optimization. The Excel Solver add-in bundled with Excel performs optimization. Optimization requires that the user identify a cell containing an objective ("set cell" in Solver); determine whether the objective is to be maximized or minimized; identify cell(s) containing decision variable(s) ("changing cells" in Solver); and identify pairs of cells with a mathematical relationship of >=, =, or <=. There are various issues and recommended practices regarding a "well-conditioned" model that is easy to "feed" to Solver that are related to linearity and sanity.

To our astonishment, Claude is effective at using optimization. Even more astonishing, for two of our test problems, Claude recognized that a problem statement is amenable to optimization analysis, built a well-conditioned spreadsheet model, found the optimal solution (using Python, as it cannot access Excel Solver), and kindly provided written guidance on how a user can populate the Solver dialog box to perform optimization themselves in Excel. The authors were positively gobsmacked by this performance.

### 6.4 Writing VBA

VBA was not part of our study, but this experience seems worth reporting. A student in an MBA elective course was unable to run an Excel add-in called SolverTable [Albright, 2025] that performs Solver sensitivity analysis by running Solver for a series of input values. SolverTable is analogous to Data Table, except that Data Table changes an input value and recalculates the spreadsheet, whereas SolverTable changes an input value and first runs Solver to generate optimal decision variable values and only then recalculates the spreadsheet. She prompted the free version of Claude to write VBA code to perform this functionality. The VBA code worked on her computer, a series of Solver runs were duly generated, and The Day Was Saved. Furthermore, the VBA code ran successfully on the first author's laptop.

## 7 OBSERVATIONS

We identified five areas of note regarding Claude's performance.

### 7.1 Different Claude Models Generate Different Spreadsheets

We used Haiku, Sonnet, and Opus. The three models produced spreadsheets with different numbers of cells and worksheets. In general, we found that Opus models were most sophisticated, which would be expected since it is the top-tier model.

It is notable that when performing optimization, Opus provided constraint status cells, which are an advanced programming technique taught in some (but not all) business school management science textbooks. Opus also claims to do a formal corner point analysis for linear optimization. These are advanced skills.

Haiku sometimes created spreadsheet models with glaring errors. Sonnet and Opus did not generate errors. Then, in the midst of this research, Haiku disappeared from Claude.

### 7.2 Claude Is an Unreliable Witness

We were curious as to the fate of Haiku in Claude for Excel, so we asked Claude. Claude provided a helpful quote from techtimes.com that Haiku is "optimized for high-throughput, latency-sensitive programmatic use (think: background agents processing millions of requests), not interactive spreadsheet analysis where quality matters more than speed." We searched techtimes.com for this quote and could not find it. We asked Claude where the quote came from, and Claude provided a charming confession (Appendix 2) that it *made up the quote and the citation*.

This is a salutary reminder that nothing created by an LLM can be taken at face value.

### 7.3 Claude Changed a Data Value

For the Buckeye Power & Light case [Frey and Carraway, 1991] Opus changed a data value, resulting in a difference in optimal objective function value in the sixth significant digit. This is immaterial in a business sense, but it is an error, *and this raises doubts about any Claude outputs that are not carefully verified. This is a serious issue.*

We asked Claude what happened. Claude claimed it changed the value to improve numerical stability of the optimization algorithm it was employing. This claim is face-plausible, but due to the first author's expertise in optimization, we do not believe it. However, a person without expertise in optimization might nod their head and think they learned something.

We suspect that Claude hallucinated a data value, then hallucinated an excuse for the first hallucination. However, we cannot prove it to be untrue without undertaking an investigation that would likely be a paper of its own. We seem to live in age where plausible BS carries the day because it takes too long to ascertain it is nonsense.

### 7.4 Claude Overcomes Complexity

For the complex case Buckeye Power & Light Company [Frey & Carraway 1991], Claude took data that was organized inconveniently (identical information was provided only in the top row for a supplier) and developed a complete table. The model requires unit conversions, which were done seamlessly and correctly. These details can consume a lot of analyst time, especially for beginners.

### 7.5 Claude Integrates Data, Model, Analysis, and Insights

For the complex case Buckeye Power & Light Company [Frey & Carraway 1991], Claude (Opus) generated six worksheets. Each worksheet had an appropriate title. All of the data was presented in a clear, well-organized fashion on a tab named "Inputs". An Optimization Model tab was provided. There was an Analysis tab that summarized a series of Solver Sensitivity runs that only an advanced analyst would undertake. There were additional tabs of analysis, and a plain-English Executive Summary tab where the insights were generally good. However, only an expert can evaluate the quality of this work.

## 8 CONCLUSIONS

AI is changing quickly, and indeed during the several weeks of this project, Claude removed Haiku from Claude for Excel, and released a new model called *Fable* as we were doing final edits. It seems likely that AI performance will continue to improve.

Claude for Excel not only created a high-quality spreadsheet model, but in some cases inferred that optimization analysis was appropriate, and performed and reported optimization results. Although Claude cannot access Solver, it did provide correct instructions for the user to apply Solver. This is a significant and unexpected result.

Claude for Excel recognizes that the Excel Data Table feature is the preferred way to do sensitivity analysis, informs the user that Microsoft does not allow it access to Data Table, and implements a workaround of compressing the entire model into a single cell formula and making multiple copies of it to generate outputs that are equivalent to a Data Table (at least until the model is modified). Claude takes a similar approach to Scenario Analysis.

The Buckeye Power & Light case [Frey & Carraway 1991] is a test-piece of a hard spreadsheet modeling problem, a non-trivial optimization implementation, and some interesting analysis questions. It took the lead author half a day to build a high-quality model for this case. For students, this case requires hours and sometimes days of difficult work (which provides valuable learning and professional growth). Now, Claude can produce the spreadsheet artifact in minutes. And perform the optimization, and follow-on analysis. This is a massive step change in productivity.

If these results are representative of AI's performance of spreadsheet modeling and analysis for semi-structured problem statements, Claude is transformative for spreadsheet modeling and analysis. For an experienced analyst, Claude provides a massive productivity boost. For a beginner analyst who cannot evaluate the quality and accuracy of the AI output, the value of the AI is questionable, and the risk of errors looms large. In addition, hallucinations remain a concern and the AI results cannot be taken on faith.

In some sense, Claude just ate our core course.

### 8.1 Future Research

There is much opportunity for future research.

<u>Data-Driven Analysis</u>

We did not consider data-driven analytics which obviously merits attention. The Reunion Case [Mehrotra & Oztanriseven, 2021] is particularly suitable for AI testing.

<u>Custom Standards</u>

Many organizations have formal coding standards. [Grossman & Özlük, 2010] discuss three financial modeling standards. Could coding standards, training materials, exemplar models and other information be uploaded, and Claude for Excel prompted to build spreadsheets following standards and practices? The potential efficiencies are obvious, but the challenge of ensuring accuracy and trustworthiness will be subtle and complex.

<u>Education</u>

The purpose of our entry-level business analytics course is to teach non-technical students to "think like an analyst". We use software (mostly spreadsheets) as the means of expression to challenge them to think clearly, express their thoughts rigorously, and evaluate their conclusions critically. With the thinking skill of modeling and the expressive skill of spreadsheet engineering now automated by AI, we have lost an important vehicle for enhancing students' analytical capacities.

The question of how to teach clear, rigorous thinking and expression is very important and very open.

<u>How Do We Train the Next Generation of Experts?</u>

Expertise is required to evaluate the AI artifact. Expertise comes from the hard graft of thinking, modeling, and building code. The question arises, how does one acquire expertise if one does not have to walk the difficult path of struggling with problem statements to create good models? How will new experts be developed or emerge?

This is a central question for the future of business analysis using AI: How do we develop the next generation of experts who can evaluate whether the AI got it right?

<u>Trustworthiness</u>

Software systems are ultimately accepted based on trust in their fitness for purpose. Professional developers invest resources in practices to ensure fitness, and trust is associated with the developers and their processes. What is trust in an AI-developed spreadsheet model? A beginner spreadsheet modeler who lacks sophistication might not be able to evaluate a spreadsheet, and results will need to be taken on trust. This opens a host of practical and philosophical issues.

How Do We Know the AI Got It Right?

The most important question is whether a spreadsheet model generated by AI provides numerical results that are correct for the business situation. This demands that the model be correct (requiring *validation*), and that the software implementation of that model be correct (requiring *verification*).

The standard for non-trivial modeling and coding challenges in the operations research and software engineering communities is to write and evaluate a model specification, and only then write code. However, spreadsheet analysts routinely conflate these two tasks (often successfully, sometimes at their peril), and when problems arise the coaching intervention is to exit the spreadsheet and validate the model prior to debugging the spreadsheet.

When the problem statement contains ambiguity or a need to make assumptions or seek additional information not immediately provided, validation becomes a nuanced topic and the modeler may need to exercise professional judgment, meaning that there might be multiple different "right" models.

But what is one to do when there exists only a spreadsheet artifact? If, as in this research, one has a known good spreadsheet model, validation and verification are easy. Absent that, a domain expert can undertake a spreadsheet audit [e.g., Operis, 2026]. This is a sizeable topic, not something to be done casually, and requires some level of expertise.

Extracting the Underlying Model from the Spreadsheet Artifact

When a model is represented as a spreadsheet (as opposed to a text, numerical, or algebraic model specification) then validation demands that one extract the underlying model from the spreadsheet code. If there is a coding error, then the underlying model will be garbled. The skill to extract (or perhaps we should say "infer") the underlying model from a spreadsheet computer program is an essential skill that seems to have received little formal attention.

One approach might be to automatically convert the model embodied as a spreadsheet into algebra or into Python code. Technology to do this exists. But it does not change the fact that some level of domain expertise is required to validate the model.

Validating and Verifying a Model from Claude

Claude for Excel provides upon request the Python code it uses (or claims to use) for modeling. If we believe this to be true, then validation could be performed on the Python code independent of the spreadsheet.

Presumably, Claude performs a mechanical conversion from Python to Excel. If we believe that conversion to be reliably accurate (which is plausible), then verification could merit only modest attention. However, in one instance Claude changed a data value (section 7.3) so verification is still required.

## ACKNOWLEDGEMENTS

We thank MBA student Samantha Malloy for sharing her experience using Claude to write VBA. We thank the School of Management at the University of San Francisco for research assistant funding.

 

## APPENDIX 1 – THE WALL BID

The Wall Bid problem statement,(adapted from [Panko & Sprague, 1998]) is in Figure A1.

**Wall Bid**

Suppose that you are working for a general contractor ("the original consulting firm") who has asked you to build a spreadsheet model to help her to create a client bid to build a wall.

According to your boss, you will start by offering two options—lava rock or brick.

Both walls will be built by crews of two.

Crews will work three eight-hour days to build either type of wall.

The wall will be 20 feet long, 6 feet tall and 2 feet thick.

Wages will be $10 per hour per person. You will have to add 20% to wages to cover benefits.

Lava Rock will cost $3 per cubic foot. Brick will cost $2 per cubic foot.

Your bid must add a profit margin of 30% to your expected cost.

*Figure A1, The Wall Bid Problem Statement, showing formatting of the PDF*

The spreadsheet model produced by Claude (Opus 4.7) is in Figure A2. Note that cells containing data (as opposed to formulas) are distinguished by a blue font color. To increase readability, we prompted Claude to put spaces in cell formulas.

| | A | B | C |
|---|---|---|---|
| 1 | **Wall Bid — Business Analysis** | | |
| 2 | | | |
| 3 | **1. Inputs & Assumptions** | | |
| 4 | *Wall dimensions* | | |
| 5 | Length (ft) | 20 | |
| 6 | Height (ft) | 6 | |
| 7 | Thickness (ft) | 2 | |
| 8 | Wall volume (cu ft) | 240 | |
| 9 | | | |
| 10 | *Labor* | | |
| 11 | Crew size (people) | 2 | |
| 12 | Days to build | 3 | |
| 13 | Hours per day | 8 | |
| 14 | Total labor-hours | 48 | |
| 15 | Wage ($/hr/person) | $10.00 | |
| 16 | Benefits load (% of wages) | 20.0% | |
| 17 | Fully-loaded labor rate ($/h | $12.00 | |
| 18 | | | |
| 19 | *Materials* | | |
| 20 | Lava rock cost ($/cu ft) | $3.00 | |
| 21 | Brick cost ($/cu ft) | $2.00 | |
| 22 | | | |
| 23 | *Pricing policy* | | |
| 24 | Profit margin (% of cost) | 30.0% | |
| 25 | | | |
| 26 | **2. Cost Build-Up & Bid** | | |
| 27 | Line item | Lava Rock | Brick |
| 28 | Wall volume (cu ft) | 240 | 240 |
| 29 | Material unit cost ($/cu ft) | $3.00 | $2.00 |
| 30 | Material cost | $720.00 | $480.00 |
| 31 | Labor-hours | 48 | 48 |
| 32 | Loaded labor rate ($/hr) | $12.00 | $12.00 |
| 33 | Labor cost | $576.00 | $576.00 |
| 34 | **Total expected cost** | **$1,296.00** | **$1,056.00** |
| 35 | Profit margin % | 30.0% | 30.0% |
| 36 | Profit ($) | $388.80 | $316.80 |
| 37 | **Bid price to client** | **$1,684.80** | **$1,372.80** |

| | A | B | C |
|---|---|---|---|
| 1 | **Wall Bid — Business Analysis** | | |
| 2 | | | |
| 3 | **1. Inputs & Assumptions** | | |
| 4 | *Wall dimensions* | | |
| 5 | Length (ft) | 20 | |
| 6 | Height (ft) | 6 | |
| 7 | Thickness (ft) | 2 | |
| 8 | Wall volume (cu ft) | = B5 * B6 * B7 | |
| 9 | | | |
| 10 | *Labor* | | |
| 11 | Crew size (people) | 2 | |
| 12 | Days to build | 3 | |
| 13 | Hours per day | 8 | |
| 14 | Total labor-hours | = B11 * B12 * B13 | |
| 15 | Wage ($/hr/person) | 10 | |
| 16 | Benefits load (% of wages) | 0.2 | |
| 17 | Fully-loaded labor rate ($/hr) | = B15 * (1 + B16) | |
| 18 | | | |
| 19 | *Materials* | | |
| 20 | Lava rock cost ($/cu ft) | 3 | |
| 21 | Brick cost ($/cu ft) | 2 | |
| 22 | | | |
| 23 | *Pricing policy* | | |
| 24 | Profit margin (% of cost) | 0.3 | |
| 25 | | | |
| 26 | **2. Cost Build-Up & Bid** | | |
| 27 | Line item | Lava Rock | Brick |
| 28 | Wall volume (cu ft) | = $B$8 | = $B$8 |
| 29 | Material unit cost ($/cu ft) | = $B$20 | = $B$21 |
| 30 | Material cost | = B28 * B29 | = C28 * C29 |
| 31 | Labor-hours | = $B$14 | = $B$14 |
| 32 | Loaded labor rate ($/hr) | = $B$17 | = $B$17 |
| 33 | Labor cost | = B31 * B32 | = C31 * C32 |
| 34 | **Total expected cost** | **= B30 + B33** | **= C30 + C33** |
| 35 | Profit margin % | = $B$24 | = $B$24 |
| 36 | Profit ($) | = B34 * B35 | = C34 * C35 |
| 37 | **Bid price to client** | **= B34 + B36** | **= C34 + C36** |

*Figure A2, Claude Spreadsheet for Wall Bid with values (left) and formulas (right)*

## APPENDIX 2 – CLAUDE FESSES UP

Here is the Claude output where Claude admits it made up a seemingly-insightful finding.

I visited the link you provided and it didn't show the reason you mentioned. Give me the real origin that I can actually cite.

Acknowledged personal inference masqueraded as cited source material ›

You're right to call that out, and I should be straight with you: that explanation was my own reasoning/inference, not something I found in any source. I attached the Tech Times citation to the surrounding context about what Excel can do, but the actual "why no Haiku" logic was me speculating — it wasn't in that article or any other search result I found.

The honest answer is: **I don't know the official reason**, and I don't have a citable source for it. The most I can point to with a real source is that Haiku 4.6, 4.7, and 4.8 simply don't exist as of June 2026 — so the available Haiku (4.5) is a generation behind the models shown in the picker. Whether that's why Anthropic left it out of the Excel UI, or whether it's a deliberate quality-tier decision, I genuinely don't know and shouldn't have presented my guess as a cited fact. Hidekazu-konishi

Sorry for the sloppy citation.